\documentclass[nofootinbib,aip,twocolumn,showpacs,amsmath,amssymb,floatfix,superscriptaddress]{revtex4}
\usepackage{graphicx}
\usepackage{color}
\graphicspath{ {./} } 

\begin{document}

\title{A new spatio-temporal description of long-delayed systems: ruling the dynamics}

\date{\today}

\author{Francesco~Marino}
\affiliation{CNR - Istituto Nazionale di Ottica, 
                largo E. Fermi 6, I-50125 Firenze, Italy}

\author{Giovanni~Giacomelli}
\affiliation{CNR - Istituto dei Sistemi Complessi, 
                via Madonna del Piano 10, I-50019 Sesto Fiorentino, Italy}

\begin{abstract}
The data generated by long-delayed dynamical systems can be organized in patterns by means of the so-called spatio-temporal representation, uncovering the role of multiple time-scales as independent degrees of freedom. However, their identification as equivalent space and time variables does not lead to a correct dynamical rule. We introduce a new framework for a proper description of the dynamics in the thermodynamic limit, providing a general avenue for the treatment of long-delayed systems in terms of partial differential equations. Such scheme is generic and does not depend on the vicinity of a super-critical bifurcation as required in previous approaches. We discuss the general validity and limit of this method and consider the exemplary cases of long-delayed excitable, bistable and Landau systems. 
\end{abstract}

\maketitle

{\it Introduction.} In spite of their long history, time-delayed systems are still an active research topic at the interface of physics, biology, mathematics and engineering \cite{Chaos2017}. Indeed, time lags appear naturally in realistic
models of disparate phenomena, e.g. whenever the finite propagation times and response-speeds or memory effects become relevant. In this context, a quite remarkable case is represented by long-delayed systems, where the delay in a feedback loop is much larger than any other characteristic time-scale involved (for a recent review, see \cite{Yanchuk2017}). The main tool for such an investigation was first introduced in \cite{Arecchi1992}, as a re-organization of the data making evident strong similarities to a one-dimensional, spatially extended system. This method, called Spatio-Temporal Representation (STR) is based on the idea that the dynamics on a single delay-cell evolves along a pseudo-time represented by the index of the subsequent cells. As such, the time variable $t$ is written as
\begin{equation}
t = \sigma +\theta T~,
\label{STR}
\end{equation} 

where $\{\sigma,\theta\}$ are named {\it pseudo-space} and {\it -time} respectively and $T$ is the delay time. 
While this mapping is always feasible, it is only in the long-delay limit that $\sigma$ and $\theta$ are well-separated timescales thus behaving, to a certain extent, as mutually independent variables \cite{Yanchuk2017}. 
In this case, a variety of equivalent spatio-temporal phenomena, hidden in the long-delayed dynamics were indeed demostrated. These include domain coarsening and nucleation \cite{Giacomelli2012,Javaloyes2015,Giacomelli2013}, front pinning and localized structures \cite{Marino2014-2017,garbin2015,Romeira2016,Marinochaos}, chimera states \cite {Chimeras} and more recently critical phase transitions \cite{Faggian}. 
In all the above situations, the identification of $\sigma$ and $\theta$ with a spatial and temporal variable respectively appeared as the most natural. Indeed, the patterns are seen to evolve over the unbounded 
$\theta$-direction, spreading through the $\sigma$-axis in a finite cell subject to (almost) periodic boundary conditions. However, such an identification cannot be easily inferred by a microscopical observation of the system (i.e., far from boundaries).

In this work, we critically discuss the STR and provide evidence that in fact it is not the appropriate framework for a spatio-temporal interpretation of the long-delay dynamics. In particular, even if the data re-organization provided by Eq. (\ref{STR}) discloses the existence of two-dimensional correlations and pattern structures, we show that an alternative setup represents the proper spatio-temporal rule in the thermodynamic limit $T \to \infty$.  

{\it Representations.} As a starting point, we recall that Eq. (\ref{STR}) must be accompanied by a suitable definition of the boundary conditions (BCs). Without loss of generality, we consider the following model  
\begin{equation}
\dot{y}(t) = G(y(t),y(t-T));
\label{del-model}
\end{equation} 

more complicated situations involving multiples variables and/or hierarchically long delays \cite{Yanchuk2014-2015,Brunner2018} can be treated in the same way. To solve Eq.(\ref{del-model}), the function $y$ must be assigned in the interval $[-T,0]$. Using (\ref{STR}) and defining $Y(\sigma,\theta) = y(t)$, the problem (\ref{del-model}) rewrites as
\begin{equation}
\partial_\sigma Y(\sigma,\theta) = G(Y(\sigma,\theta),Y(\sigma,\theta-1))~,
\label{str-model}
\end{equation} 

and the initial value problem translates into
\begin{equation}
Y(\sigma,-1) =  \phi(\sigma)~,~\sigma \in (0,1] ~,
\label{str-initial}
\end{equation}

with the BC
\begin{equation}
Y(\sigma +T,\theta -1) =  Y(\sigma,\theta) ~,
\label{str-bc}
\end{equation}

where $\phi$ is $Y$ as assigned in the first delay interval. 

Notably, in the presence of correlations of $y$ in consecutive delay units, the pattern in the space $(\sigma,\theta)$ is correlated along the $\theta$ direction. In this case, the condition (\ref{str-bc}) can be written in the thermodynamic limit as
\begin{equation}
Y(\sigma+T,\theta) \approx  Y(\sigma,\theta) ~,
\label{str-bc2}
\end{equation}

in analogy with the periodic BCs used in spatially-extended (SE) systems. 

The STR framework is thus defined by Eqs. (\ref{str-model}), (\ref{str-initial}) and (\ref{str-bc2}), which lead to the commonly adopted identification of $\sigma$ and $\theta$ as pseudo-space and -time respectively. This idea is supported by the behavior of the maximal comoving Lyapunov exponent \cite{Giacomelli1996,Giacomelli1998}, which also yields a clear determination of the intrinsic drift present in long-delayed system due to causality \cite{Yanchuk2017}.

\begin{figure}
\begin{center}
\includegraphics*[width=0.9\columnwidth]{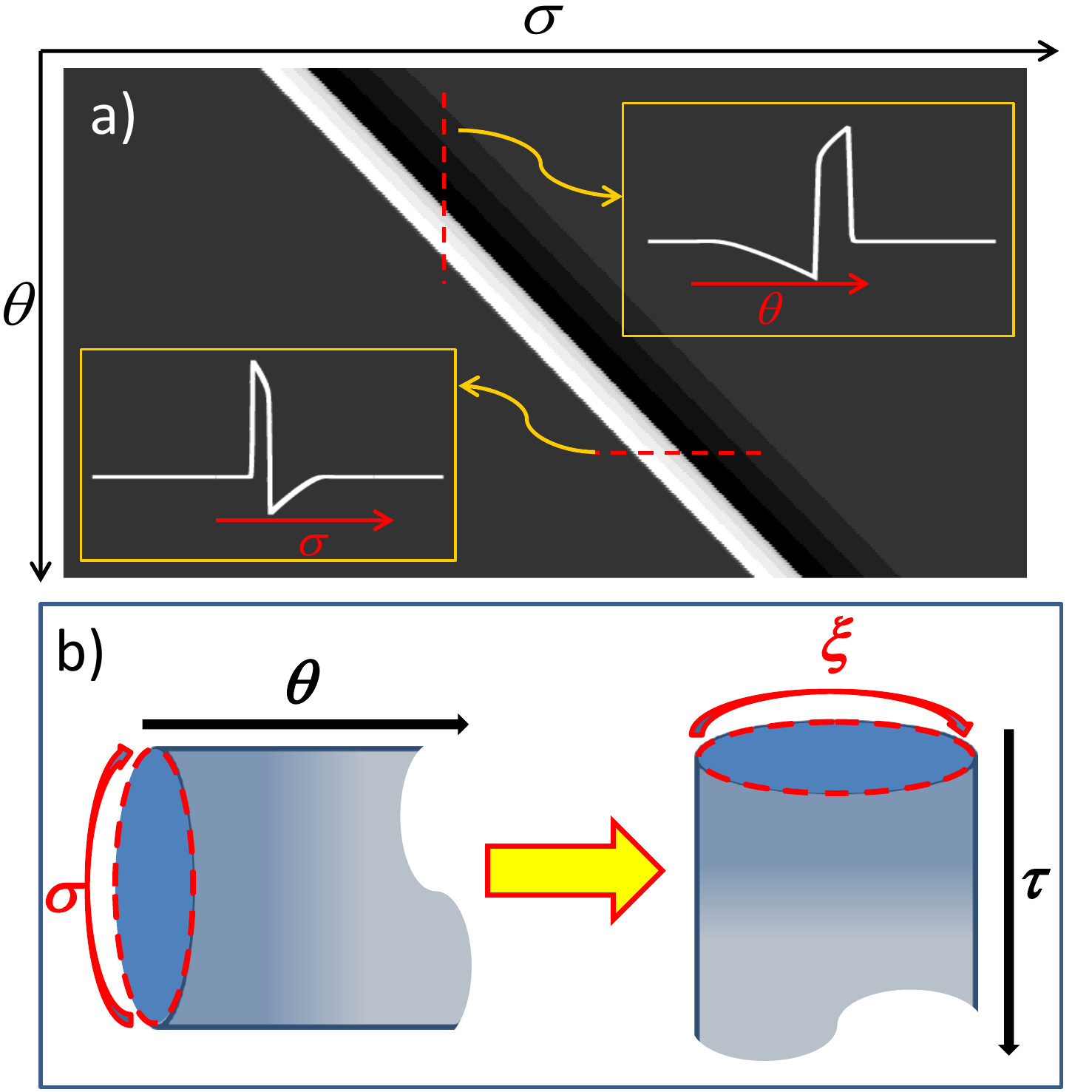}
\end{center} 
\caption{a) Excitable pulse propagation in the space-time $\{\sigma,\theta\}$ obtained by numerical integration of Eq. (\ref{fhn-d}) with zero noise. Parameters: $g=$ 0.1, $\tau=2\times10^3$, $J=$~-0.51, $\alpha=$1.5, $\varepsilon$=$0.01$. The insets represent two transverse cuts along $\sigma$ and $\theta$ directions. b) Pictorial view of the manifolds associated to the STR (left) and DR (right). The dashed circular lines mark the initial condition domains. The curved and straight arrows indicate respectively the periodic BCs and the direction of evolution.
\label{fig1}}
\end{figure}

On the other hand, Eq.(\ref{str-model}) does not provide an explicit evolution rule in $\theta$. The only method to derive it has been a multiple-scale approach  \cite{Giacomelli1996,Kashchenko1998,Yanchuk2014-2015,Wolfrum2006,Giacomelli1998,Bestehorn2000}, separating fast and slow scales into different perturbation orders close to a super-critical bifurcation. Nevertheless, the generalization of the above scheme to other cases, e.g. involving finite amplitude solutions is not straightforward. One of these situations is represented by the long-delayed Fitz-Hugh Nagumo (FHN) system, introduced in \cite{Marinochaos} to model an excitable semiconductor laser with feedback,
\begin{eqnarray}
\dot{u} &=& F(u) +w +g u(t-T) +\zeta \nonumber\\
\dot{w} &=& -\varepsilon~(w - J + \alpha u)~,
\label{fhn-d}
\end{eqnarray}
 
that describes the evolution of two variables $\{u,w\}$ evolving with characteristic time-scales whose ratio is the small parameter $\varepsilon$. Here the function $F(u) = u - u^3$ describes the polarization dynamics, $g$ is the feedback gain, $J$ the pump current, $\alpha$ a coupling coefficient and $\zeta$ is a $\delta$-correlated, white Gaussian noise. 
In Eqs.(\ref{fhn-d}), an in-homogeneous initial condition or sufficiently strong perturbation triggers the emission of excitable pulses that propagate in the pseudo-spacetime $\{\sigma,\theta\}$. The evolution of one of these pulses is shown 
in Fig. \ref{fig1}a. After some transient, the pulse propagates with a constant velocity and a fixed shape that is independent on the initial conditions: these features are immediately reminiscent of what is observed in 1D spatially-extended excitable systems (see e.g. \cite{Murray}). Nevertheless, the pattern here observed displays a peculiar aspect which is inherently related to the STR of the delayed dynamics. The refractory tail, i.e. the slow, negative recovery of the quiescent state in response to a perturbation {\it anticipates} the excited state along the $\theta$ direction. Such a paradoxical behavior where an effect actually precedes the cause is not consistent with the idea that $\theta$ is the genuine time variable. In fact, this role appears to be more properly embodied by $\sigma$, since along its direction the refractory tail {\it follows} the pulse as expected (see the insets of  Fig. \ref{fig1}a). 

On the basis of the above observations,  we postulate that the correct rule for generating the equivalent spatio-temporal evolution of a long-delayed system in the limit $T \to \infty$ should consider $\{\theta,\sigma\}$ as space- and time- variable respectively. 

We name such new description Dynamical Representation (DR), and we denote the corresponding space- and time- variables as $ \{\xi,\tau\}$ in place of $\{\theta,\sigma\}$. Defining $Z(\xi,\tau) = Y(\sigma,\theta)$, Eq.(\ref{str-model}) rewrites as the explicit evolution rule
\begin{equation}
\partial_\tau Z(\xi,\tau) = G(Z(\xi,\tau),Z(\xi-1,\tau))~.
\label{DR}
\end{equation} 

The delayed term now becomes a non-local asymmetric, spatial coupling which breaks the $\xi$-spatial symmetry. In the following, we will consider spatially-periodic BCs
\begin{equation}
Z(\xi+S,\tau) = Z(\xi,\tau) ~,
\label{DR-bc}
\end{equation}
 
with a size $S = [ t_{tot}/T ]$, where $[.]$ denotes the integer part and $t_{tot}$ is the total time span. 

The domains associated to the STR and DR are depicted in Fig. \ref{fig1}b, evidencing different global manifolds: the dashed circular lines mark the initial conditions, the cylinder axis defines the direction of evolution (time-axis) and the cross-sectional circumference corresponds to the size of the spatial cell. 

The physical meaning of the DR can be enlighten considering the following example. For the linear, long delayed system
\begin{equation}
\dot{y}(t) = \mu y(t) +g y(t-T),
\label{lin-del}
\end{equation} 

rewritten in the STR as
\begin{equation}
\partial_\sigma Y(\sigma,\theta) = \mu Y(\sigma,\theta) +g Y(\sigma,\theta-1),
\label{lin-del-STR}
\end{equation} 

the solution can be obtained in the Laplace domain as
\begin{equation}
\chi(\bar{\sigma},\bar{\theta}) = {1 \over \bar{\sigma} -\mu - g e^{-\bar{\theta}}},
\label{lin-del-model}
\end{equation} 

where $\{\bar{\sigma},\bar{\theta}\} \subset \mathbb{C}$ are the Laplace-conjugate variables of  $\{\sigma,\theta\}$. $\chi$ can be interpreted as the response to a stimulus and must satisfy the Kramers-Kr\"onig relations to obey causality \cite{toll}. One can readily verify that this is actually the case when considering the variable $\bar{\sigma}$, but it is not with $\bar{\theta}$. 

The above example indicates that already in this simple situation, while the STR provides a suitable method to build a meaningful reorganization of the data in a pattern, it cannot be used straightforwardly to generate a genuine
spatiotemporal dynamics, i.e. satisfying the causality.  

{\it Formal expansion.} The DR provides an explicit spatio-temporal rule, but a more useful description can be obtained from an equivalent Partial-Differential Equation (PDE) model. This can be pursued by formally expanding the non-local term as
\begin{equation}
Z(\xi-1,\tau) \approx Z(\xi,\tau) -Z_\xi(\xi,\tau) +{1\over 2}Z_{\xi\xi}(\xi,\tau)-..~,
\label{expansion}
\end{equation}

where $Z_\xi = \partial_\xi Z,~ Z_{\xi\xi} = \partial^2_{\xi\xi} Z, ..$, obtaining the PDE
\begin{equation}
Z_\tau = G(Z,Z_\xi,Z_{\xi\xi},..)~.
\label{pde}
\end{equation}

The validity of the expansion (\ref{expansion}) relies on the a-posteriori examination of the dynamics generated by Eq.(\ref{del-model}), since the scale of the evolution along $\xi$ cannot be generally determined in advance. 
However, in the absence of an anomalous Lyapunov exponent \cite{Giacomelli1995} (or in the weak-chaos limit \cite{Heiligenthal2011}) the correlation along $\xi$ decays over a length $L_\xi \gg 1$. Upon rescaling $\xi \to \xi/L_\xi$, the convergence of (\ref{expansion}) can be made explicit with a smallness parameter $1/L_\xi \ll 1$. In these conditions, the applicability of Eq. (\ref{expansion}) relies on the smoothness of the pattern solution and thus should not depend on its amplitude or the vicinity to a bifurcation. We will show that this is indeed the case in the examples described below.

Depending on the system and the order of the expansion, the $\xi$-spatial symmetry-breaking induced by the non-local coupling may be included or not in the resulting model. Here we consider the case of a linear delayed term only, leaving for a future work a more general discussion. In this class of models, each order of the expansion adds a specific feature: 
the $0^{th}$ is a re-normalization of the local force, the $1^{st}$ provides the linear drift (that can be removed with a suitable choice of a co-moving reference frame), the $2^{nd}$ the linear diffusion, the $3^{rd}$ the first non-trivial spatial symmetry-breaking term, etc. While not all the orders of the expansion lead to a numerically stable model, in general the dynamics of Eq. \ref{DR} is better approximated by including increasingly higher-order terms. We finally point out that the coefficients of different orders share the gain factor in the original expression and thus are not independent.

{\it Delayed FHN.} In the regime where localized structures are solutions of (\ref{fhn-d}), the expansion (\ref{expansion}) can be performed for the corresponding DR, obtaining at the second order  
\begin{eqnarray}
U_\tau &=& F(U) +W +g U -g U_\xi  +{g\over 2} U_{\xi\xi}+\zeta  \nonumber \\
W_\tau &=& -\varepsilon~(W - J + \alpha U)~,
\label{fhn-exp}
\end{eqnarray}

where $\{U,W\} = \{U(\xi,\tau),W(\xi,\tau)\} = \{u(t),w(t)\}$.

\begin{figure}
\begin{center}
\includegraphics*[width=1.0\columnwidth]{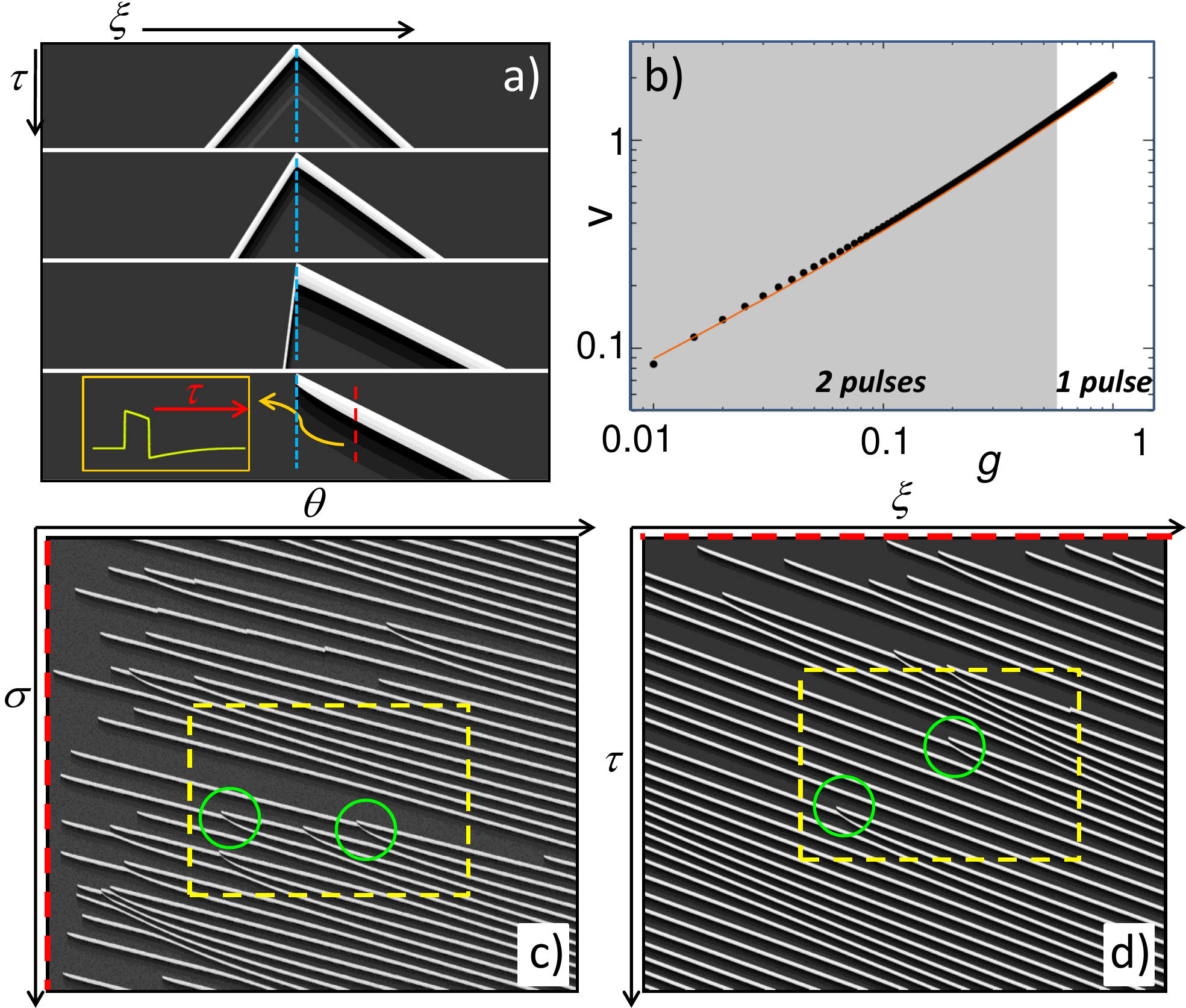}
\end{center} 
\caption{a) Propagation of excitable pulses from the numerical integration of Eq. (\ref{fhn-exp}). From top to bottom: $g=$ 0.01, $g=$ 0.1, $g=$ 0.56, $g=$ 0.6. Other parameters: $J=$~-0.51, $\alpha=$1.5, $\varepsilon$=$0.01$, noise amplitude $3\times 10^{-2}$. The size of the ($\xi$, $\tau$) space-time cell is 1600 $\times$ 400.  The inset is a cut along the $\tau$ direction. b) propagation velocities of excitable pulses obtained by numerical integration of Eq.(\ref{fhn-d}) (symbols) and Eq.(\ref{fhn-exp}) (solid line). c,d): spatiotemporal plots of noise-induced excitable pulses from Eq. (\ref{fhn-d}) (c) and Eq. (\ref{fhn-exp}) (d). In both cases the size of the space-time cells is 4000 $\times$ 8000. As in Fig. \ref{fig1}b, the red dashed lines on the vertical (c) and horizontal (d) axis depict the initial conditions domain.
\label{fig2}}
\end{figure}

The above equations represent the well-known FHN model with advection \cite{kneer}. Simulations of (\ref{fhn-exp}) are presented in Fig.\ref{fig2}a, with a narrow initial condition to trigger the excitable response. As seen in the panels, for low values of the gain $g$ two excitable pulses are generated with an asymmetry both in shape and propagation. Increasing the gain, the difference between the pulses increases up to the disappearance of the second one. Notably, the $2^{nd}$ order expansion does not breaks the spatial symmetry since the first order spatial derivative can be removed with the choice of a co-moving reference frame. However, there exists a balance between the advection term and the diffusion such that for high $g$ the second pulse is suppressed (bottom panel) \cite{zykov,meron}. In the original system (\ref{fhn-d}), only a single pulse is always observed, confirming that the additional symmetry-breaking induced by the full non-local terms suppresses the second pulse. 
In this sense our system is more similar to 1D spatially-extended FHN model with strong advection. 
The situation depicted in the bottom panel of Fig.\ref{fig2}a is indeed very close to the findings in (\ref{fhn-d}) (see Fig.\ref{fig1}a). This observation can be quantified by measuring the pulse velocity as a function of the gain; the results are plotted in Fig.\ref{fig2}b. It is seen that, even in the regimes where two pulses are present in the system (\ref{fhn-exp}), the velocity of the first pulse is in a good agreement with that of the solitary pulse found in (\ref{fhn-d}), confirming the validity of the expansion approach.

In Fig.\ref{fig2}c-d, we compare the patterns obtained from the STR of Eq. (\ref{fhn-d}) and those obtained from (\ref{fhn-exp}). In the presence of noise both systems displays the sporadic emission of excitable pulses. With exception of the initial transients, in the bulk the two patterns are remarkably similar (see the yellow dashed boxes). In particular, the interaction events (green circles in the middle of Figs.\ref{fig2}c-d) display the very same features, where one of the lowest of two neighboring pulses is deviated and starts following a downward-curved trajectory. This is due to the fact that in the DR the refractory tail of each pulse, which is responsible of such repulsive interaction \cite{garbin2015,Marinochaos}, always appear below the excited region. 

{\it Delayed bistable.} As seen in the former example, the DR expansion allows us to describe long-delayed systems in terms of PDEs, even in regimes where finite-amplitude solutions occurs. In this context, another important case is represented by the long-delayed, bistable system introduced as a phenomenological model for a bistable semiconductor laser with feedback \cite{Giacomelli2012}
\begin{equation}
\dot{y}(t) = F(y(t)) +g y(t-T)~,
\label{bis-d}
\end{equation} 

where now $F(y) = -y(y-1)(y+1+a)$ is a force derived from a quartic potential characterized by an asymmetry $a$ and $g$ is the feedback gain. The above model has been succesfully applied to describe several phenomena such as the generation, propagation and annihilation of quasi-heteroclinic fronts, nucleation and coarsening \cite{Giacomelli2012, Giacomelli2013}.

In the DR, Eq.(\ref{bis-d})  writes
\begin{equation}
\partial_\tau Z(\xi,\tau) = F(Z(\xi,\tau)) +g Z(\xi-1,\tau)~,
\label{bis-DR}
\end{equation} 

and, expanding up to the second order we obtain
\begin{equation}
\partial_\tau Z = F(Z) +g Z -gZ_\xi +{g\over 2}Z_{\xi\xi}~,
\label{bis-exp}
\end{equation} 

i.e. a reaction-diffusion process with advection, characterized by a drift velocity $v_d=g$, a diffusion $ D = {1\over 2}g$ and a new effective force $\bar F(Z) = F(Z) +g Z$. In this model the velocity of the fronts can be computed analytically \cite{Murray} obtaining $c_\pm = g \pm {a \over 2} g^{1/2}$ that coincides with the estimation reported in \cite{Giacomelli2012} for Eq. (\ref{bis-d}). 

We point out that (\ref{bis-exp}) is trasversally-symmetric in the comoving reference frame of the advection term. As a consequence, specific symmetry-breaking phenomena such as the asymmetric annihilation of fronts observed in the long-delayed system \cite{Giacomelli2013} cannot occur. These could be recovered by adding suitable (odd) higher order terms in the expanded model (\ref{bis-exp}). 

\begin{figure}
\begin{center}
\includegraphics*[width=1.0\columnwidth]{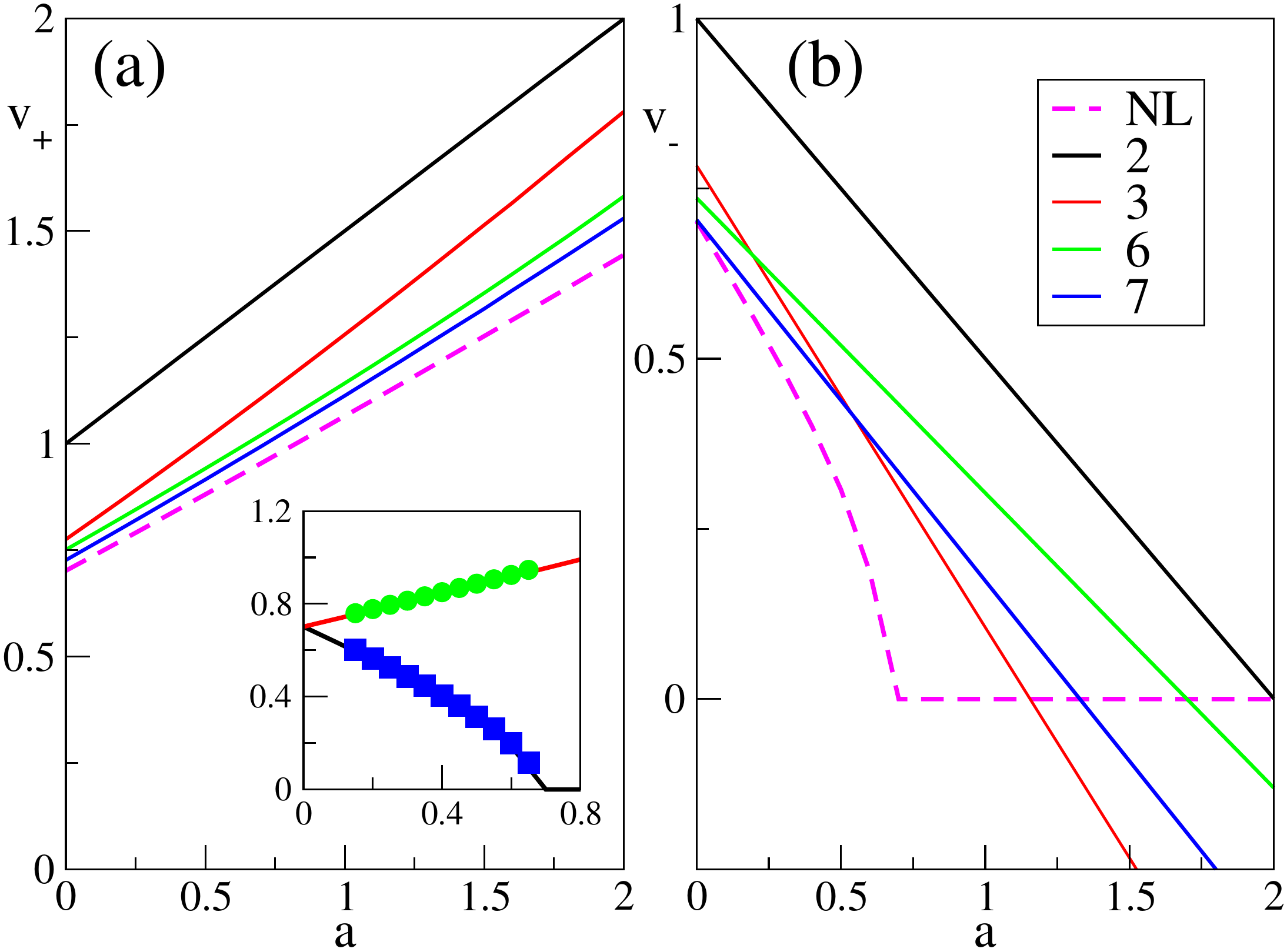}
\end{center} 
\caption{Front velocities in the bistable model for $g=1$ as a function of the asymmetry, for an increasing order of the expansion. Dashed: result for model (\ref{bis-DR}). Inset: comparison between (\ref{bis-DR}) (lines) and (\ref{bis-d}) (dots and squares).
\label{bis}}
\end{figure}

In Fig.\ref{bis} we report the numerical estimation of the fronts velocities for increasing order of the expansion, comparing with those of (\ref{bis-DR});  in the inset, it is reported the direct comparison of the velocities evaluated from model (\ref{bis-DR}) and the original (\ref{bis-d}). As shown in the figure, $v_+$ is better estimated increasing the order as expected. A more complicated situation appears for $v_-$. For the model (\ref{bis-DR}), the velocity drops and remains equal to zero beyond a certain asymmetry value, while the velocities for the different expansion orders decrease monotonically (with some crossings between the orders which are still under investigation). This behavior can be understood by considering that a front cannot propagate backwards in $\xi$ due to the non-local coupling, while such a bound does not hold for the PDE models, in a close analogy to what found in the delayed and spatial FHN system about the existence of the second pulse. 

{\it Delayed complex Landau.} To our knowledge, few setups exist where a mapping between a long-delayed dynamical system and a PDE has been established. We mention the Delayed Complex Landau (DCL) model both for a single \cite{Giacomelli1996,Kashchenko1998} and two hierarchical \cite{Yanchuk2014-2015} long delays, in the case of Eckhaus instability \citep{Wolfrum2006}, and the rate equation model of a class-B laser with feedback \cite{Giacomelli1998,Bestehorn2000}. In all the above studies, the analysis was performed with a multiple-scale method in the vicinity of a super-critical Hopf bifurcation. To compare our approach with the above results, we consider the DCL model
\begin{equation}
{\dot y}(t) = \mu y(t) -(1+i \beta)|y(t)|^2 y(t) +g y(t-T)~,
\label{dcl}
\end{equation} 

where $y$ is complex, and we write the DR description
\begin{equation}
Z_\tau = \mu Z -(1+i \beta)|Z|^2 Z +g Z(\xi-1,\tau)~.
\label{dcl-DR}
\end{equation}

We begin our comparison by noting that the maximal comoving Lyapunov exponent can be computed analytically (with the proper BCs) in the linear case for this model as well, and it coincides with that reported in \cite{Giacomelli1996}. 

The 2$^{nd}$ order expansion of (\ref{dcl-DR}) writes as
\begin{equation}
Z_\tau = (\mu +g) Z -gZ_\xi +{g\over 2} Z_{\xi\xi} -(1+i \beta)|Z|^2 Z~,
\label{dcl-exp}
\end{equation}

e.g. a Complex Ginzburg-Landau (CGL) equation with drift $g$ and diffusion ${g\over 2}$. This coincides with the findings of \cite{Giacomelli1996} and \citep{Wolfrum2006} (for their coefficient $\beta=0$) after a suitable coordinate exchange. 

\begin{figure}
\begin{center}
\includegraphics*[width=1\columnwidth]{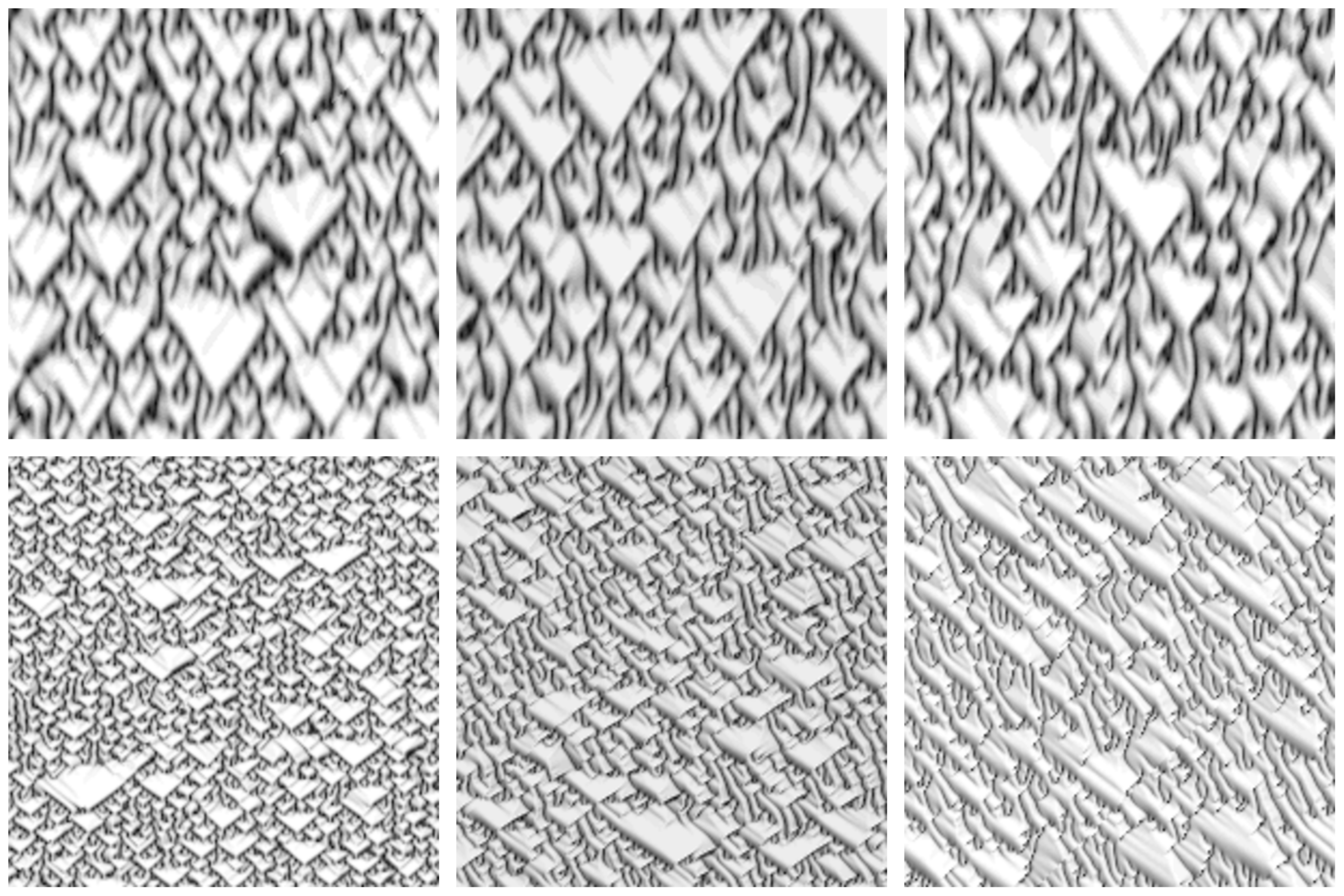}
\end{center} 
\caption{Simulation of the $2^{nd}$ order (model (\ref{dcl-exp}), left column) and $3^{rd}$ order expansion (central column) of (\ref{dcl-DR}). Right column: simulation of model (\ref{dcl}). All results for $\beta=3$ and $g=1$, shown in the comoving frame. Top row: $\mu = -0.8$ (close to the Hopf bifurcation $\mu = -1$), bottom row: $\mu=1$.}
\label{dcl-cgl}
\end{figure}

Model (\ref{dcl-DR}) gives a very good description of the bulk dynamics of (\ref{dcl}), while model (\ref{dcl-exp}) should instead represent a valid approximation only close the Hopf bifurcation at $\mu_H = -g$. Indeed, this is the case as shown in Fig. \ref{dcl-cgl}. Increasing $\mu$, we move away from the Hopf bifurcation and strong spatial asymmetries appear in the simulations of (\ref{dcl}).
This features cannot be reproduced by the spatially-symmetric (in the comoving frame) model (\ref{dcl-exp}) as indeed shown in the figure, where for the higher $\mu$ the pattern is still spatially symmetric. To deal with this, we added the next order (the third) in the expansion of (\ref{dcl-DR}). As an odd-order, we expect to obtain a spatial symmetry-breaking which can fit more closely the simulation of the system (\ref{dcl}). This is indeed what we found, with a similar behavior close to to the Hopf bifurcation but with a better approximation of model  (\ref{dcl}) far from it. Higher orders further improve the approximation and will be discussed elsewhere.

{\it Conclusions.}  We have introduced and discussed an alternative approach to the spatio-temporal re-organization of data generated from a long-delayed dynamical system. In this framework, the bulk dynamics is produced with a new rule, employing the opposite definition of equivalent space and time variables with respect to the STR. While the domain manifolds for the two methods are quite different (the bound and unbound variables are exchanged), we have shown that the bulk dynamics (away from the boundaries, or equivalently in the thermodynamic limit) is more properly obtained in the new representation. We expect that this new rule should not change the statistical properties of the generated patterns as measured by auto-correlations and the Kolmogorov-Sinai entropy \cite{Giacomelli1995}, since they are expressed as bulk properties as well. The method also allows for a straightforward expansion of the non-local coupling in term of spatial derivatives, leading eventually to a normal form description through standard PDEs.

{\it Acknowledgments.} We thank A. Torcini for useful hints about the time-split integration method.

\end{document}